\documentclass[oneside,11pt]{article}

\usepackage[utf8]{inputenc}
\usepackage[T1]{fontenc}

\usepackage{amsthm}
\usepackage{amsmath}
\usepackage{amsfonts}
\usepackage{amssymb}

\usepackage{graphicx} 
\usepackage{float}
\usepackage{booktabs}
\usepackage{multirow}
\usepackage{rotating}
\usepackage{array}
\usepackage{xcolor}
\usepackage{subfig}
\usepackage[ruled]{algorithm2e}

\usepackage{breakcites}

\usepackage{enumitem}

\usepackage[top=1.5cm,bottom=2cm,right=2.5cm,left=2.5cm]{geometry}
\usepackage{titling}

\usepackage{natbib}

\usepackage{ifplatform}

\usepackage{textcomp}

\ifwindows
\fi

\ifwindows
\usepackage{bbm}
\else
\iflinux
\usepackage{bbm}
\else
\usepackage{bbm}
\fi
\fi

\DeclareFontFamily{OT1}{pzc}{}
\DeclareFontShape{OT1}{pzc}{m}{it}{<-> s * [1.10] pzcmi7t}{}
\DeclareMathAlphabet{\mathpzc}{OT1}{pzc}{m}{it}

\usepackage[pdftex,final,bookmarksnumbered,bookmarksopen=false,breaklinks,colorlinks]{hyperref}
\hypersetup{
	final,
	bookmarksnumbered=true,
	bookmarksopen=true,
	bookmarksopenlevel=0,
	unicode=false,          
	pdftoolbar=true,        
	pdfmenubar=true,        
	pdffitwindow=false,     
	pdftitle={Goodness-of-fit tests for functional linear models based on integrated projections},    
	pdfdisplaydoctitle=true, 
	pdftoolbar=true,
	pdfmenubar=true,
	pdflang={English},
	pdfauthor={Eduardo Garcia-Portugues, Javier Alvarez-Liebana, Gonzalo Alvarez-Perez, Wenceslao Gonzalez-Manteiga},     
	pdfsubject={arXiv paper},   
	pdfcreator={Eduardo Garcia-Portugues},   
	pdfproducer={Eduardo Garcia-Portugues}, 
	pdfkeywords={Functional data}{Graphical tool}{Projections}{Regularization}, 
	pdfnewwindow=true,      
	breaklinks=true,
	hidelinks,       
	linkcolor=black,          
	citecolor=black,        
	filecolor=magenta,      
	urlcolor=cyan           
}

\newtheorem{lemma}{Lemma}

\allowdisplaybreaks

\newif\ifmain
\maintrue
\ifmain

\else

\fi
\newif\ifsupplement
\supplementtrue
\newif\iffigstabs
\figstabstrue

\begin{document}

\ifmain

\title{Goodness-of-fit tests for functional linear models based on integrated projections}
\setlength{\droptitle}{-1cm}
\predate{}%
\postdate{}%

\date{}

\author{Eduardo Garc\'ia-Portugu\'es$^{1,5}$, Javier \'Alvarez-Li\'ebana$^{2}$, Gonzalo \'Alvarez-P\'erez$^{3}$,\\ and Wenceslao Gonz\'alez-Manteiga$^{4}$}

\footnotetext[1]{Department of Statistics, Carlos III University of Madrid (Spain).}
\footnotetext[2]{Department of Statistics and Operations Research and Mathematics Didactics, University of Oviedo (Spain).}
\footnotetext[3]{Department of Physics, University of Oviedo (Spain).}
\footnotetext[4]{Department of Statistics, Mathematical Analysis and Optimization, University of Santiago de Compostela (Spain).}
\footnotetext[5]{Corresponding author. e-mail: \href{mailto:edgarcia@est-econ.uc3m.es}{edgarcia@est-econ.uc3m.es}.}

\maketitle


\begin{abstract}
	Functional linear models are one of the most fundamental tools to assess the relation between two random variables of a functional or scalar nature. This contribution proposes a goodness-of-fit test for the functional linear model with functional response that neatly adapts to functional/scalar responses/predictors. In particular, the new goodness-of-fit test extends a previous proposal for scalar response. The test statistic is based on a convenient regularized estimator, is easy to compute, and is calibrated through an efficient bootstrap resampling. A graphical diagnostic tool, useful to visualize the deviations from the model, is introduced and illustrated with a novel data application. The R package \texttt{goffda} implements the proposed methods and allows for the reproducibility of the data application.
\end{abstract}
\begin{flushleft}
\end{flushleft}

\section{Functional linear models}

\subsection{Formulation}

Given two separable Hilbert spaces $\mathbb{H}_1$ and $\mathbb{H}_2$, we consider the regression setting with centered $\mathbb{H}_2$-valued response $\mathcal{Y}$ and centered $\mathbb{H}_1$-valued predictor $\mathcal{X}$:
\begin{align}
\mathcal{Y}=m(\mathcal{X})+\mathcal{E},\label{eq:modreg}
\end{align}
where $m:{\scriptstyle\mathcal{X}}\in\mathbb{H}_1\mapsto \mathbb{E} \left[ \mathcal{Y}|\mathcal{X}={\scriptstyle{\mathcal{X}}} \right]\in\mathbb{H}_2$ is the regression operator and the $\mathbb{H}_2$-valued error $\mathcal{E}$ is such that $\mathbb{E} \left[ \mathcal{E} | \mathcal{X} \right] = 0$. When $\mathbb{H}_1 = L^2 \left([a,b] \right)$ and $\mathbb{H}_2 = L^2 \left([c,d]\right)$, the Functional Linear Model with Functional Response (FLMFR; see, e.g., \citet[Chapter 16]{RamsaySilverman05}) is the most well-known parametric instance of \eqref{eq:modreg}. If the regression operator is assumed to be Hilbert--Schmidt, $m$ is parametrizable as 
\begin{align}
m_{\beta} (\mathcal{X})=\int_{a}^{b} \beta(s,\cdot)\mathcal{X}(s)\,\mathrm{d}s=:\langle\langle\beta, \mathcal{X}\rangle\rangle,\label{eq:lin}
\end{align}
for $\beta\in\mathbb{H}_1\otimes\mathbb{H}_2=L^2\left([a,b]\times[c,d]\right)$ a square-integrable kernel. The present work considers this framework and is concerned with the goodness-of-fit of the family of $\mathbb{H}_2$-valued and $\mathbb{H}_1$-conditioned linear models
\begin{align}
\mathcal{L} := \left\lbrace  \langle \langle \beta,\cdot \rangle \rangle:\beta\in \mathbb{H}_1\otimes\mathbb{H}_2\right\rbrace.\label{eq:L}
\end{align}

Any $\mathcal{X} \in \mathbb{H}_1$ and $\mathcal{Y},\mathcal{E} \in \mathbb{H}_2$ can be represented in terms of orthonormal bases $\lbrace \Psi_j \rbrace_{j=1}^{\infty}$ and $\lbrace \Phi_k \rbrace_{k=1}^{\infty}$ as $\mathcal{X} = \sum_{j=1}^{\infty} x_j \Psi_j$, $\mathcal{Y} = \sum_{k=1}^{\infty} y_k \Phi_k$, and $\mathcal{E} = \sum_{k=1}^{\infty} e_k \Phi_k$, where $x_j = \langle \mathcal{X}, \Psi_j \rangle_{\mathbb{H}_1}$, $y_k = \langle \mathcal{Y}, \Phi_k \rangle_{\mathbb{H}_2}$, and $e_k = \langle \mathcal{E}, \Phi_k \rangle_{\mathbb{H}_2}$, $\forall j,k \geq 1$. Also, $\beta\in\mathbb{H}_1\otimes\mathbb{H}_2$ can be expressed as
\begin{align*}
\beta = \sum_{j=1}^\infty \sum_{k=1}^\infty b_{jk} (\Psi_j \otimes \Phi_k),\quad b_{jk} = \left\langle \beta, \Psi_j\otimes\Phi_k\right\rangle_{\mathbb{H}_1\otimes\mathbb{H}_2},\quad \forall j,k \geq 1.
\end{align*}
Therefore, the population version of the FLMFR based on \eqref{eq:lin} can be expressed as 
\begin{align}
y_k = \sum_{j=1}^{\infty} b_{j k}x_{j} + e_k, \ k \geq 1.\label{eq:inf}
\end{align}

\subsection{Model estimation}
\label{sec:est}

The projection of \eqref{eq:inf} into the truncated bases $\lbrace \Psi_{j}\rbrace_{j=1}^p$ and $\lbrace \Phi_{k}\rbrace_{k=1}^q$ opens the way for the estimation of $\beta$ given a centered sample $\{(\mathcal{X}_{i},\mathcal{Y}_{i})\}_{i=1}^n$. Indeed, the truncated sample version of \eqref{eq:inf} is expressed\nolinebreak[4] as
\begin{align}
\mathbf{Y}_{q} = \mathbf{X}_{p}  \mathbf{B}_{p,q}+\mathbf{E}_q,\label{eq:mult}
\end{align}
where $\mathbf{Y}_{q}$ and $\mathbf{E}_{q}$ are $n\times q$ matrices with the respective coefficients of $\{\mathcal{Y}_i\}_{i=1}^n$ and $\{\mathcal{E}_i\}_{i=1}^n$ on $\lbrace \Phi_{k}\rbrace_{k=1}^q$, $\mathbf{X}_{p}$ is the $n\times p$ matrix of coefficients of $\{\mathcal{X}_i\}_{i=1}^n$ on $\lbrace \Psi_{j}\rbrace_{j=1}^p$, and $\mathbf{B}_{p,q}$ is the $p\times q$ matrix of coefficients of $\beta$ on $\lbrace \Psi_{j}\otimes\Phi_k\rbrace_{j,k=1}^{p,q}$.\\

Several estimators for $\beta$ have been proposed; see, e.g., \cite{Yaoetal05}, \cite{Heetal10}, \cite{CrambesMas13}, \cite{Benatiaetal17}, and \cite{ImaizumiKato18}. A popular estimation paradigm is Functional Principal Components Regression (FPCR; \cite{RamsaySilverman05}), which considers the (empirical) Functional Principal Components (FPC) $\lbrace \hat\Psi_{j}\rbrace_{j=1}^p$ and $\lbrace \hat\Phi_{k}\rbrace_{k=1}^q$ as a plug-in for $\lbrace \Psi_{j}\rbrace_{j=1}^p$ and $\lbrace \Phi_{k}\rbrace_{k=1}^q$ underneath \eqref{eq:mult}. Estimation by FPCR yields $\hat{\mathbf{B}}_{p,q}=\mathrm{arg}\min_{\mathbf{B}_{p,q}}\left\| \mathbf{Y}_q-\mathbf{X}_{p}  \mathbf{B}_{p,q} \right\|^{2} \allowbreak= \big(\mathbf{X}_{p}' \mathbf{X}_{p} \big)^{-1} \mathbf{X}_{p}' \mathbf{Y}_{q}$, with $j=1,\ldots,p$ and $k=1,\ldots,q$. The estimator $\hat{\mathbf{B}}_{p,q}$ depends on $(p,q)$ and an automatic data-driven selection of $(p,q)$ is of most practical interest. However, cross-validatory procedures are computationally expensive, especially since two tuning parameters must be optimized. A simple alternative for selecting $q$ is to guarantee a certain proportion of explained variance (say, $0.99$) for $\{\mathcal{Y}_{i}\}_{i=1}^n$. The more critical selection of $p$ can be done by first ensuring a certain proportion of explained variance (say, $0.99$) and then performing a LASSO-regularized FPCR regression (FPCR-L1 henceforth):
\begin{align*}
\hat{\mathbf{B}}_{p,q}^{(\lambda)}=\mathrm{arg}\min_{\mathbf{B}_{p,q}}\left\{\frac{1}{2n}  \sum_{i=1}^{n} \left\|\left(\mathbf{Y}_q \right)_i - \left(\mathbf{X}_{p}  \mathbf{B}_{p,q}\right)_i
\right\|^{2} + \lambda \sum_{j=1}^{p} \left\|\left(\mathbf{B}_{p,q}\right)_j\right\|
\right\},
\end{align*}
where the notation $(\mathbf{A})_i$ stands for the $i$-th row of the matrix $\mathbf{A}$. This regularization applies a row-wise penalty that enables variable selection for a given $\lambda$, which can be efficiently selected by cross-validation and its \emph{one standard error} variant \citep{Friedman2010}. \\

However, FPCR-L1 lacks an explicit expression for the hat matrix (in contrast with FPCR), an important handicap for the bootstrap algorithm outlined in Section \ref{sec:boot}. To combine the flexible variable selection of FPCR-L1 with the analytical form of FPCR, we propose the FPCR-L1S estimator, which firstly implements FPCR-L1 for variable selection and then performs FPCR on the selected predictors. It returns the hat matrix $\mathbf{H}_\mathrm{C}^{(\lambda)} = \tilde{\mathbf{X}}_{\tilde{p}} \big(\tilde{\mathbf{X}}_{\tilde{p}}' \tilde{\mathbf{X}}_{\tilde{p}}\big)^{-1} \tilde{\mathbf{X}}_{\tilde{p}}'$, where $\tilde{\mathbf{X}}_{\tilde{p}}$ is the matrix of the coefficients of the $\tilde{p}$ LASSO-selected predictors (not necessarily sorted).\\

Simulations \citep[Section 2.4]{Garcia-Portugues:arXiv} report that FPCR-L1S outperforms FPCR.

\section{Proposed goodness-of-fit tests}

\subsection{Test statistic genesis}

Our aim is to test whether the regression operator belongs to the class of linear operators described in \eqref{eq:L}, that is, to test
\begin{align*}
\mathcal{H}_0:m \in \mathcal{L}\quad\text{vs.}\quad \mathcal{H}_1:  m \not\in \mathcal{L}.
\end{align*}

To do so, we use the following lemma to characterize $\mathcal{H}_0$ in terms of the one-dimensional projections of  $\mathcal{Y}$ and $\mathcal{X}$. The lemma requires from analogues of the Euclidean $(p-1)$-sphere $\mathbb{S}^{p-1}:=\{\mathbf{x}\in\mathbb{R}^{p}:\|\mathbf{x}\|=1\}$: the $(p-1)$-sphere of $\mathbb{H}_1$ for $\lbrace \Psi_{j}\rbrace_{j=1}^\infty$, $\mathbb{S}_{\mathbb{H}_1,\lbrace \Psi_{j}\rbrace_{j=1}^\infty}^{p-1} := \lbrace \sum_{j=1}^{p}x_{j}\Psi_{j}\in \mathbb{H}_1: \left\| \mathbf{x}\right\|=1 \rbrace$ and, analogously, $\mathbb{S}_{\mathbb{H}_2,\{\Phi_k\}_{k=1}^\infty}^{q-1}$.

\begin{lemma}[$\mathcal{H}_0$ characterization on finite-dimensional directions; \cite{Garcia-Portugues:arXiv}\label{lemma:1}]
Let $\mathcal{X}$ and $\mathcal{Y}$ be $\mathbb{H}_1$- and $\mathbb{H}_2$-valued random variables, respectively, $\beta \in \mathbb{H}_{1}\otimes \mathbb{H}_{2}$, and let $\{\Psi_j\}_{j=1}^\infty$ and $\{\Phi_k\}_{k=1}^\infty$ be bases of $\mathbb{H}_1$ and $\mathbb{H}_2$, respectively. Then, the next statements are equivalent:
\begin{enumerate}[label=\roman*., ref=\textit{\roman*}]
	\item $\mathcal{H}_0$ holds, that is, $m \left({\scriptstyle\mathcal{X}} \right) = \langle \langle {\scriptstyle\mathcal{X}}, \beta \rangle \rangle$, $\forall {\scriptstyle\mathcal{X}} \in \mathbb{H}_1$.\label{lemma:1:1}
	\item $\mathbb{E} \Big[ \big\langle \mathcal{Y} - \langle \langle \mathcal{X}, \beta \rangle \rangle, \gamma_{\mathcal{Y}}^{(q)} \big\rangle_{\mathbb{H}_2} \mathbbm{1}_{\big\{\big\langle \mathcal{X}, \gamma_{\mathcal{X}}^{(p)} \big\rangle_{\mathbb{H}_1} \leq u\big\}}\Big] = 0$, for almost every $u \in \mathbb{R}$, $\forall \gamma_{\mathcal{X}}^{(p)} \in \mathbb{S}_{\mathbb{H}_1,\{\Psi_j\}_{j=1}^\infty}^{p-1},\, \forall\gamma^{(q)}_{\mathcal{Y}} \in \mathbb{S}_{\mathbb{H}_2,\{\Phi_k\}_{k=1}^\infty}^{q-1}$, and for all $p,q \geq 1$. \label{lemma:1:8}
\end{enumerate}
\end{lemma}
The reader is referred to \cite{Garcia-Portugues:arXiv} for the proof of the lemma.\\

We use the above characterization to detect deviations from $\mathcal{H}_0$. We do so by means of the $(p,q)$-truncated empirical version of the doubly-projected integrated regression function in statement \ref{lemma:1:8}, that is, the residual marked empirical process
\begin{align}
R_{n,p,q} \big(u, \gamma_{\mathcal{X}}^{(p)},\gamma_{\mathcal{Y}}^{(q)} \big) = \frac{1}{\sqrt{n}}
\sum_{i=1}^{n} \big\langle \hat{\mathcal{E}}_{i}^{(q)}, \gamma_{\mathcal{Y}}^{(q)} \big\rangle_{\mathbb{H}_2} \mathbbm{1}_{\big\lbrace \big\langle \mathcal{X}_{i}^{(p)}, \gamma_{\mathcal{X}}^{(p)}\big\rangle_{\mathbb{H}_1} \leq u \big\rbrace},\quad u \in \mathbb{R},\label{eq:proc}
\end{align}
with residual marks $\big\langle \hat{\mathcal{E}}_{i}^{(q)}, \gamma_{\mathcal{Y}}^{(q)} \big\rangle_{\mathbb{H}_2} = \hat{\mathbf{e}}_{i,q}' \mathbf{h}_q$ and jumps $\big\langle \mathcal{X}_{i}^{(p)}, \gamma_{\mathcal{X}}^{(p)}\big\rangle_{\mathbb{H}_1}=\mathbf{x}_{i,p}' \mathbf{g}_p$, where $\hat{\mathbf{e}}_{i,q}'$ represents the $i$-th row of the $n\times q$ matrix of residual coefficients $\hat{\mathbf{E}}_{q}$ on $\lbrace \Phi_{k}\rbrace_{k=1}^q$, $\mathbf{x}_{i,p}$ are the first $p$ coefficients of $\mathcal{X}_i$ on $\lbrace \Psi_{j}\rbrace_{j=1}^p$, and $\mathbf{g}_p\in \mathbb{S}^{p-1}$ and $\mathbf{h}_q\in \mathbb{S}^{q-1}$ are the coefficients of $\gamma_{\mathcal{X}}^{(p)}$ and $\gamma_{\mathcal{Y}}^{(q)}$, respectively.\\

To measure the proximity of \eqref{eq:proc} to zero (and hence to $\mathcal{H}_0$), and following the ideas of \cite{Escanciano06} and \cite{GarciaPortuguesetal14}, we consider a Cram\'er--von Mises norm on $\Pi^{(p,q)} = \mathbb{S}_{\mathbb{H}_2,\{\Phi_k\}_{k=1}^\infty}^{q-1}\times \mathbb{S}_{\mathbb{H}_1,\{\Psi_j\}_{j=1}^\infty}^{p-1} \times \mathbb{R}$, yielding the so-called Projected Cram\'er--von Mises (PCvM) statistic:
\begin{align*}
\mathrm{PCvM}_{n,p,q} =\int_{ \mathbb{S}^{q-1} \times \mathbb{S}^{p-1} \times \mathbb{R} } \left[R_{n,p,q} \left(u,\mathbf{g}_p, \mathbf{h}_q \right) \right]^2 \,F_{n,\mathbf{g}_p} (\mathrm{d}u) \,\mathrm{d}\mathbf{g}_p \,\mathrm{d}\mathbf{h}_q,
\end{align*}
where $F_{n,\mathbf{g}_p}$ is the empirical cumulative distribution function of $\lbrace \mathbf{x}_{i,p}' \mathbf{g}_p\rbrace_{i=1}^n$.\\

From the developments in \cite{Garcia-Portugues:arXiv}, we get an easily computable form of the statistic:
\begin{align}
\mathrm{PCvM}_{n,p,q} = \frac{1}{n^2} \frac{2\pi^{p/2 + q/2 -1}}{q \Gamma(p/2) \Gamma(q/2)} {\rm Tr} \left[ \hat{\mathbf{E}}_{q}' \mathbf{A}_{\bullet} \hat{\mathbf{E}}_{q} \right], \label{eq:stat}
\end{align}
where ${\rm Tr}(\cdot)$ denotes the trace operator and $\mathbf{A}_{\bullet}$ is a certain $n\times n$ symmetric matrix that only depends on $\{\mathbf{x}_{i,p}\}_{i=1}^p$.

\subsection{Statistic interpretation and particular cases}

The statistic \eqref{eq:stat} can be regarded as a weighted quadratic norm:
\begin{align*}
\mathrm{PCvM}_{n,p,q} = \frac{1}{n^2} \frac{2\pi^{p/2 + q/2 -1}}{q \Gamma(p/2) \Gamma(q/2)}  \sum_{k=1}^{q} \left\| \left(\hat{e}_{1,k}, \ldots, \hat{e}_{n,k} \right)  \right\|_{\mathbf{A}_{\bullet}},
\end{align*}
where $\hat{\mathcal{E}}_{i}^{(q)} = \sum_{k=1}^{q} \hat{e}_{i,k} \Phi_{k}$, $i=1,\ldots,n$, and $\left\| \mathbf{v} \right\|_{\mathbf{A}_{\bullet}}:=(\mathbf{v}'\mathbf{A}_\bullet\mathbf{v})^{1/2}$ is a norm in $\mathbb{R}^{n}$ induced by $\mathbf{A}_{\bullet}$. Therefore, the statistic aggregates across the dimensions of the truncated response the $\mathbf{A}_\bullet$-weighted norms of the coefficients of the functional errors on $\{\Phi_k\}_{k=1}^q$. The basis of such interpretation is the next lemma (proof given in \cite{Garcia-Portugues:arXiv}).

\begin{lemma}[\cite{Garcia-Portugues:arXiv}\label{lem:4}]
Assume that the functional sample $\{\mathcal{X}_i\}_{i=1}^n$ has pairwise \emph{distinct} coefficients $\{\mathbf{x}_{i,p}\}_{i=1}^n$ on an arbitrary $p$-truncated basis $\lbrace \Psi_{j}\rbrace_{j=1}^p$ of $\mathbb{H}_1$. Then, for any sample size $n\geq1$, the $n\times n$ matrix $\mathbf{A}_\bullet$ is positive definite.
\end{lemma}

The general framework of the FLMFR seamless adapts to scalar response or predictor. So do the estimation methods discussed in Section \ref{sec:est} and the statistic \eqref{eq:stat}. Indeed, in the case of scalar response (see, e.g., \cite{Cardotetal99} and \cite{Crambesetal09}), $\mathbb{H}_2=\mathbb{R}$ is identifiable with the subspace of $L^2([c,d])$ of constant functions with basis $\{(d-c)^{-1/2}\}$ and $\beta(\cdot,\star) \equiv \beta(\cdot)\in L^2([a,b])$ is a univariate function. The statistic $\mathrm{PCvM}_{n,p,1}$ precisely corresponds to the PCvM statistic for the functional linear model with scalar response given in \cite{GarciaPortuguesetal14}. In the case of scalar predictor (see \cite{Chiouetal03}), $\beta(\cdot,\star) \equiv \beta(\star)\in L^2([c,d])$ and $\mathrm{PCvM}_{n,1,q}$ results in a test statistic specific for such model.

\subsection{Bootstrap calibration and graphical tool}
\label{sec:boot}

The calibration of the statistic \eqref{eq:stat} is done through a wild bootstrap on the residuals. We sketch next the main steps of such resampling, referring to Algorithm 1 in \cite{Garcia-Portugues:arXiv} for the specifics and its adaptation to the $\beta$-specified case.

\begin{enumerate}
\item Compute the statistic $\mathrm{PCvM}_{n,\tilde{p},q}$ from the residuals $\hat{\mathbf{e}}_{i,q} = \mathbf{Y}_{i,q} - \mathbf{X}_{i,\tilde{p}} \hat{\mathbf{B}}^{(\lambda),\mathrm{C}}_{\tilde{p},q}$, $i=1,\ldots,n$, associated to the FPCR-L1S estimate $\hat{\mathbf{B}}_{\tilde{p},q}^{(\lambda), \mathrm{C}}$ (which selects $\tilde{p}$).
\item For $b=1,\ldots,B$:
\begin{enumerate}
	\item Perturb the residuals as $\mathbf{e}^{*b}_{i,q} := V_i^{*b} \hat{\mathbf{e}}_{i,q}$, $i=1,\ldots,n$, where $\lbrace V_{i}^{\ast b}\rbrace_{i=1}^{n}$ are independent zero-mean and unit-variance random variables.
	\item Using $\{\mathbf{e}^{*b}_{i,q}\}_{i=1}^n$, simulate $\{\mathbf{Y}_{i,q}^{\ast b}\}_{i=1}^n$ from the multivariate linear model.
	\item Fit the multivariate model from $\{(\mathbf{X}_{i,\tilde{p}},\mathbf{Y}_{i,q}^{\ast b})\}_{i=1}^n$ and obtain $\hat{\mathbf{B}}_{\tilde{p},q}^{\ast b}$.
	\item Compute the bootstrapped statistic $\mathrm{PCvM}^{\ast b}_{n,\tilde{p},q}$ from the bootstrap residuals $\hat{\mathbf{e}}_{i,q}^{\ast b}:=\mathbf{Y}_{i,q}^{\ast b}-\mathbf{X}_{i,\tilde{p}}\hat{\mathbf{B}}_{\tilde{p},q}^{\ast b}$, $i=1,\ldots,n$.
\end{enumerate}
\item Estimate the $p$-value by Monte Carlo as $\#\lbrace \mathrm{PCvM}_{n,\tilde{p},q} \leq \mathrm{PCvM}_{n,\tilde{p},q}^{\ast b} \rbrace / B$.
\end{enumerate}

The bootstrap procedure yields as a by-product a graphical diagnostic tool of the goodness-of-fit of the FLMFR that helps visualizing the possible deviations from $\mathcal{H}_0$. The tool compares the empirical process on which the PCvM statistic is applied,
\begin{align*}
R_{n,p,q} \left(u,\mathbf{g}_p, \mathbf{h}_q \right) = \frac{1}{\sqrt{n}}
\sum_{i=1}^{n} \hat{\mathbf{e}}_{i,q}' \mathbf{h}_q \mathbbm{1}_{\big\lbrace \mathbf{x}_{i,p}' \mathbf{g}_p \leq u \big\rbrace},
\end{align*} 
with $G$ samples of its bootstrapped version:
\begin{align*}
R_{n,p,q}^{*b} \left(u,\mathbf{g}_p, \mathbf{h}_q \right) = \frac{1}{\sqrt{n}}
\sum_{i=1}^{n} (\hat{\mathbf{e}}_{i,q}^{*b})' \mathbf{h}_q \mathbbm{1}_{\big\lbrace \mathbf{x}_{i,p}' \mathbf{g}_p \leq u \big\rbrace},\quad b=1,\ldots,G.
\end{align*}

The graphical tool employs the FPC bases $\lbrace \hat\Psi_{j}\rbrace_{j=1}^p$ and $\lbrace \hat\Phi_{k}\rbrace_{k=1}^q$ and considers $\mathbf{g}_p$ and $\mathbf{h}_q$ as the canonical vectors in $\mathbb{R}^p$ and $\mathbb{R}^q$, respectively. This allows to visualize the deviations from $\mathcal{H}_0$ when ``it is projected'' in the first FPC of $\{\mathcal{X}_i\}_{i=1}^n$ and the first FPC of $\{\mathcal{Y}_i\}_{i=1}^n$ (or any other combination thereof). Figure \ref{fig:boot} shows and explains two outputs of this diagnostic tool, for the situations in which $\mathcal{H}_0$ is and is not rejected.

\section{Application: AEMET temperatures dataset}

The \texttt{aemet\_temp} dataset in the \texttt{goffda} \citep{Garcia-Portugues:goffda} package contains daily temperatures of $n=73$ weather stations from the Meteorological State Agency of Spain (AEMET) during the time span 1974--2013. The dataset is split in two 20-year periods, 1974--1993 and 1994--2013, and the daily temperatures on each weather station are averaged for both periods. This results in two functional samples for the average temperatures across Spain on 1974--1993 (predictor $\mathcal{X}$) and 1994--2013 (response $\mathcal{Y}$). Both samples were smoothed with local linear estimators using cross-validated bandwidths to ease visualization. Figure \ref{fig:aemet} (left) shows the samples of $\mathcal{X}$ and \nolinebreak[4]$\mathcal{Y}$.

\begin{figure}[!h]
	\vspace*{-0.25cm}
	\centering
	\includegraphics[width=0.461\textwidth]{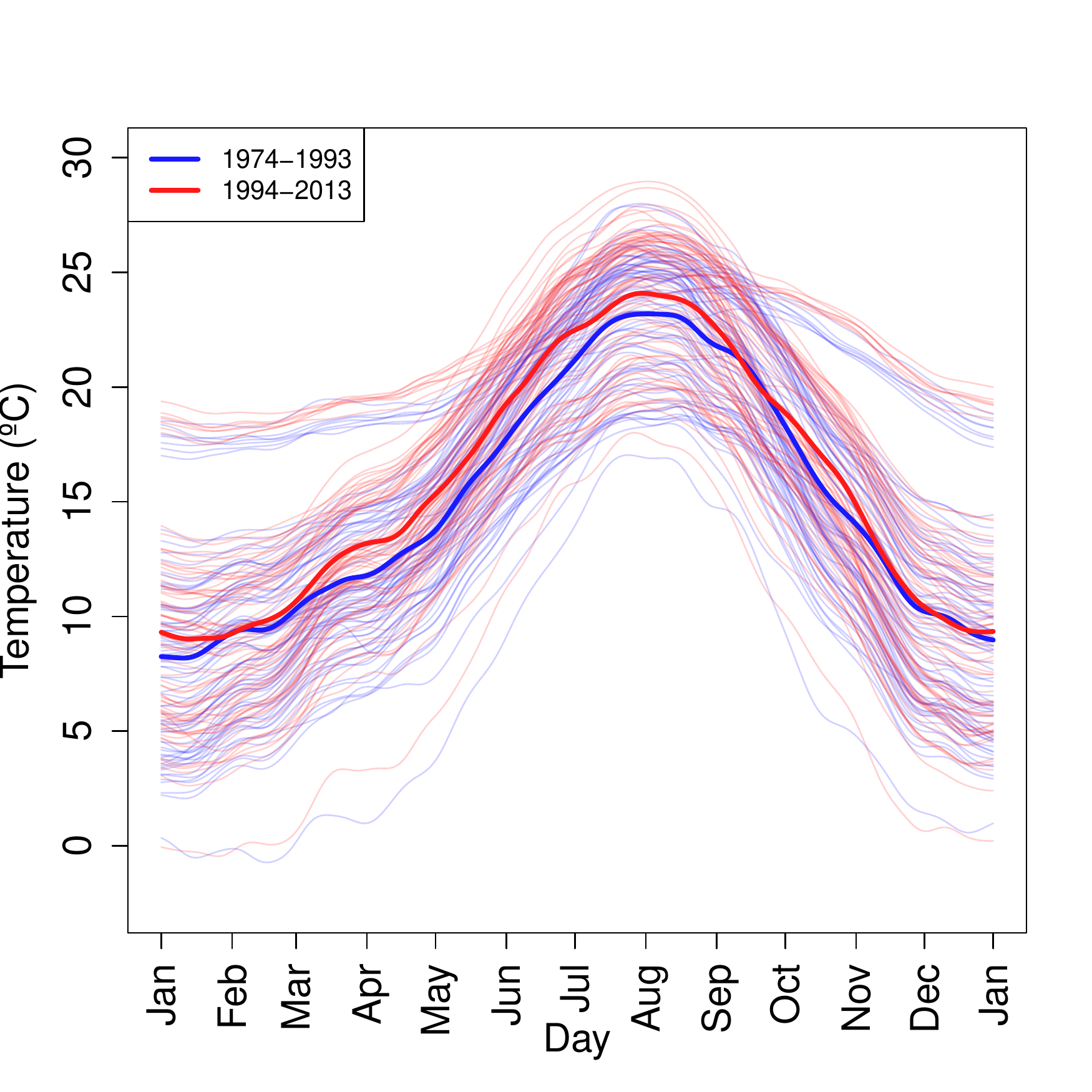}\includegraphics[width=0.5\textwidth]{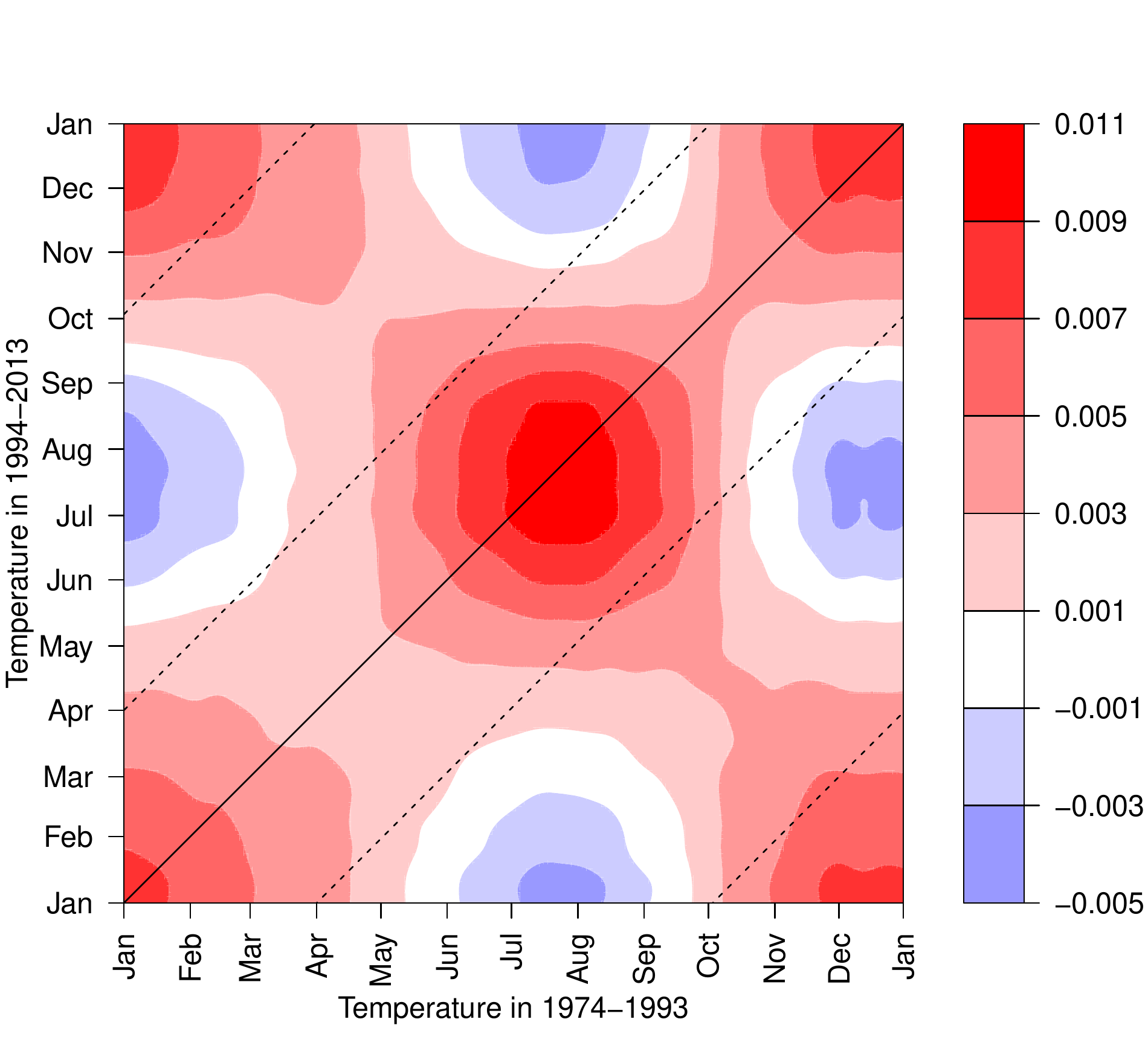}
	\caption{\small Left: Temperatures of $73$ AEMET weather stations for the periods 1974--1983 ($\mathcal{X}$) and 1994--2013 ($\mathcal{Y}$), along with their means. Right: FPCR-L1S estimator $\hat\beta$ for the FLMFR.\label{fig:aemet}}
\end{figure}

The PCvM test based on $\tilde{p}=4$ (selected by FPCR-L1S with $\lambda$ chosen by one standard error cross-validation) and $q=3$ (selected such that the proportion of explained variance is $0.99$) yielded a $p$-value equal to $0.4155$ using $B=10^4$ bootstrap replicates. Therefore, the FLMFR is not rejected. The estimated $\beta$, shown in Figure \ref{fig:aemet} (right), reveals a temperature increment on the latter period with respect to the former, a conclusion supported by the predominance of positive values on the $\hat\beta$ surface and the positiveness of almost all the temperature curves. The diagnostic tool in Figure \ref{fig:boot} (left) shows no remarkable deviations of the residual marked empirical process from $\mathcal{H}_0$. The PCvM test rejects emphatically the simple hypotheses $\mathcal{H}_0:\beta=0$ and $\mathcal{H}_0:\beta(s,t)=\mathbbm{1}_{\{s=t\}}$ (stationary-temperature hypothesis; right panel in Figure \ref{fig:boot}), thus corroborating a significant change in the temperatures between both periods. The diagnostic tool for the latter hypothesis reveals that the non-stationarity is due to the relations between the second FPC of $\{\mathcal{X}_i\}_{i=1}^n$ and $\{\mathcal{Y}_i\}_{i=1}^n$, both related with the variation shape of the temperature curves along the year.

\section{Software: \texttt{goffda} R package}

The R package \texttt{goffda} \citep{Garcia-Portugues:goffda} implements all the methods described and allows for replication of the data application. The implementation of the critical parts of the goodness-of-fit tests, such as the computation of the $\mathbf{A}_\bullet$ matrix and the computation of the PCvM statistic, are implemented in C++ (through \texttt{Rcpp} \cite{Rcpp}) for the sake of efficiency. The \texttt{goffda} package relies on the \texttt{fdata} class from the \texttt{fda.usc} \citep{Febrero-Bande2012} package, so it is fully compatible with the latter. \\

The main functions of \texttt{goffda} are: \texttt{flm\_est} (several estimation methods for the FLMFR); \texttt{Adot} (efficient implementation of the $\mathbf{A}_\bullet$ matrix); \texttt{flm\_stat} (computation of \eqref{eq:stat}); \texttt{flm\_test} (implementation of the test with its bootstrap resampling). \texttt{flm\_est} and \texttt{flm\_test} deal seamlessly with either functional/scalar responses/predictors.

\begin{figure}[h!]
	\centering
	\includegraphics[width=0.5\textwidth,trim={0cm 0cm 0.5cm 0cm},clip]{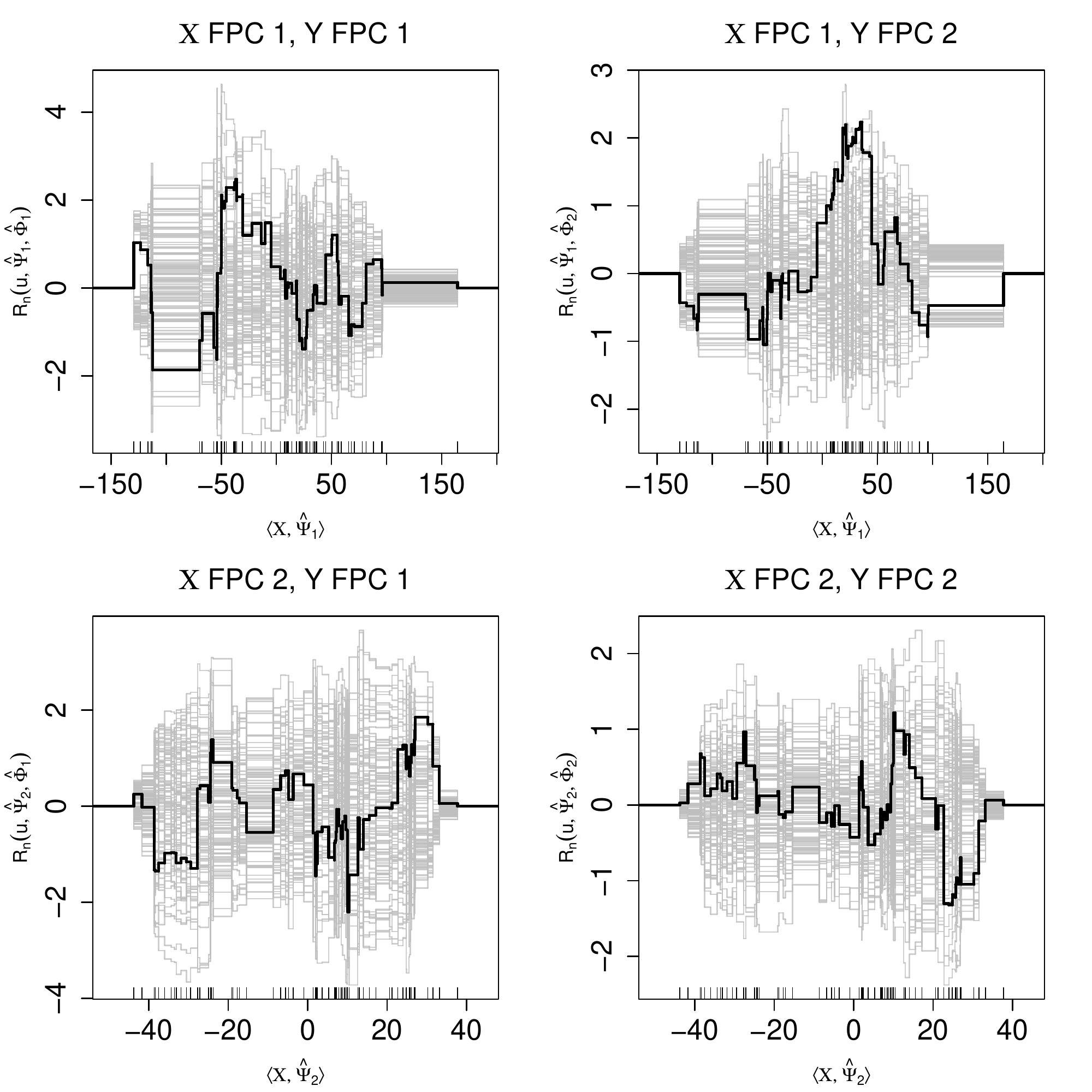}\includegraphics[width=0.5\textwidth,trim={0cm 0cm 0.5cm 0cm},clip]{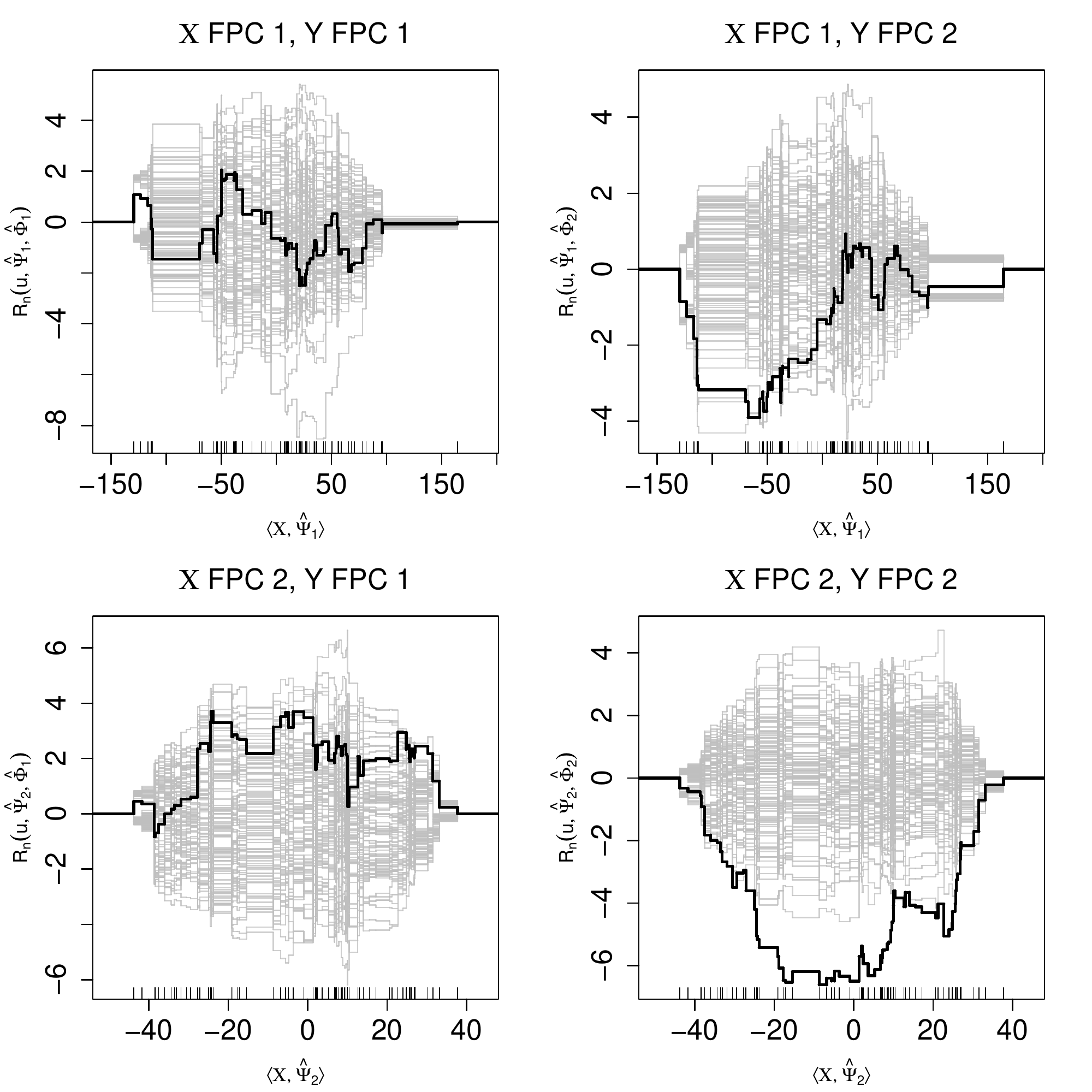}
	\caption{\small Graphical tool of the PCvM test. The black curve represents the observed process $R_{n,p,q} \left(u,\mathbf{e}_j, \mathbf{e}_k \right)$ for its projections on the $j$-th FPC of $\{\mathcal{X}_i\}_{i=1}^n$ and the $k$-th FPC of $\{\mathcal{Y}_i\}_{i=1}^n$, $j,k=1,2$. The grey curves stand for the bootstrapped processes under $\mathcal{H}_0$, i.e., $R_{n,p,q}^{*b} \left(u,\mathbf{e}_j, \mathbf{e}_k \right)$, $b=1,\ldots,100$. The left $2\times2$ panel shows the diagnostic output for $\mathcal{H}_0:m\in\mathcal{L}$ in the AEMET temperatures dataset. The non-rejection of $\mathcal{H}_0$ is manifested in the centrality of the observed process within the bootstrapped ones. The right $2\times2$ panel shows the diagnostic for $\mathcal{H}_0:\beta(s,t)=\mathbbm{1}_{\{s=t\}}$, with rejection of $\mathcal{H}_0$ evidenced by the outlyingness of $R_{n,p,q} \left(u,\mathbf{e}_2, \mathbf{e}_2 \right)$.}
	\label{fig:boot}
\end{figure}

\section*{Acknowledgements}

The authors gratefully thank Prof. Manuel Febrero-Bande for discussions and for providing access to the dataset of raw AEMET temperatures. The first author acknowledges support from grants PGC2018-097284-B-I00 and IJCI-2017-32005 from the Spanish Ministry of Economy and Competitiveness (co-funded with FEDER funds). The second author acknowledges support from grant PGC2018-099549-B-I00 from the same agency. The first and fourth authors acknowledge support from grant MTM2016-76969-P also from the same agency. The authors gratefully acknowledge the computing resources of the Supercomputing Center of Galicia (CESGA).


\end{document}